\documentclass[aps,prd,amsmath
,eqsecnum            
,preprintnumbers
,nofootinbib         
,twocolumn         
]{revtex4}
\usepackage[dvips]{graphicx}
\usepackage{amsmath}
\usepackage{amsfonts}
\usepackage{amssymb}
\usepackage{longtable}   

\allowdisplaybreaks[4]     

\newcommand{\df}{\ {\overset {\rm def} =}\ }
\newcommand{\dr}[2]{\frac {{\rm d} {#1}} {{\rm d} {#2}}}

\newcommand{\dril}[2]{{{\rm d} {#1}} / {{\rm d} {#2}}}

\begin{document}

\title{Repeatable light paths in the conformally flat cosmological models}

\author{Andrzej Krasi\'nski}
\affiliation{N. Copernicus Astronomical Centre, Polish Academy of Sciences, \\
Bartycka 18, 00 716 Warszawa, Poland} \email{akr@camk.edu.pl}

\date { }

\begin{abstract}
This is a supplement to an earlier paper (PRD{\bf 84}, 023510 (2011)), where
those shearfree normal cosmological models were identified, in which all light
rays have repeatable paths. All of them are conformally flat, but less general
than the Stephani model and more general than Robertson -- Walker. In this note,
their defining feature is identified: in each of them, in comoving coordinates,
the time-dependence factors out so that the cofactor is a static metric. An
example is given of a congruence of test observers and sources in the Minkowski
spacetime that displays nonrepeatable light paths.
\end{abstract}

\maketitle

\section{The motivation}\label{motiv}

This paper is a supplement to Ref. \cite{Kras2011}, which in turn was a 
continuation of Ref. \cite{KrBo2011}. In Ref. \cite{KrBo2011} null geodesics in 
the $\beta' \neq 0$ Szekeres models \cite{Szek1975a} -- \cite{PlKr2006} were 
investigated, and it turned out that, in general, they have nonrepeatable paths. 
This means, given a fixed comoving light source S and a fixed comoving observer 
O, two light rays emitted from S at different instants that hit O, intersect 
different sequences of matter world lines on the way. The observer will thus see 
the image of the source drift across the sky. This is a  potentially observable 
effect. It exists also for nonradial rays in the spherically symmetric 
Lema\^{\i}tre \cite{Lema1933} -- Tolman \cite{Tolm1934} model, but is 
identically zero in the Robertson -- Walker (RW) models. Thus, it could be used 
as an observational test of homogeneity of the Universe -- see an 
astronomy-oriented discussion in Refs. \cite{QQAm2009} -- \cite{QABC2012}, where 
the drift was termed ``cosmic parallax''. 

In Ref. \cite{KrBo2011} it was shown that the drift vanishes for \textit{all}
null geodesics only when the Szekeres model reduces to the Friedmann limit. The
condition for this is the same as for zero shear in the flow of the cosmic
medium. This gave rise to the question whether the drift is caused by shear or
rather by the inhomogeneity.

To clarify this question, in Ref. \cite{Kras2011} the condition for zero drift 
for all null geodesics was investigated in the shearfree normal (SFN) models 
\cite{Barn1973} -- \cite{Kras1997}. They are the solutions of Einstein's 
equations with a perfect fluid source that have zero shear, zero rotation and 
nonzero expansion in the cosmic fluid. If shear were indeed the cause of the 
drift, then in the SFN models the drift should vanish. It turned out that, in 
general, this is not the case. These models consist of 3 Petrov type D metrics 
that are spherically, plane and hyperbolically symmetric, and of the conformally 
flat Stephani solution \cite{Step1967,Kras1997} that, in general, has no 
symmetry. It was found that in each of these solutions, a drift-free subcase 
exists \cite{Kras2011} that has zero conformal curvature, but is less general 
than the zero-Weyl-tensor limit of the relevant case. At the same time, each 
subcase is more general than the RW limit, having nonzero acceleration. This 
gives rise to one more problem: what is the underlying cause of the 
repeatability of all light paths when these models are non-RW. 

This is the question answered here. It is shown that each of the drift-free
cases, in the comoving coordinates, is a conformal image of a static spacetime.
The key point is not just conformal equivalence (all these models are
conformally flat), but the form of the conformal factor. This will be explained
in Sec. \ref{characteristic}.

The repeatability of light paths (RLP) is defined relative to a family of 
observers and light sources. In a cosmological spacetime, such as Szekeres or  
Lema\^{\i}tre -- Tolman or SFN or RW, it is natural to assume the light sources 
and observers comoving with the cosmic medium, as in Refs. 
\cite{KrBo2011,Kras2011}. But one can as well assume the light sources and 
observers moving along a congruence of timelike curves unrelated to the flow 
lines of the cosmic matter, and investigate the RLP property for them. It is 
shown in the Appendix that even in the Minkowski spacetime a timelike congruence 
can be devised that displays the non-RLP property. 

\section{The drift-free SFN models}\label{SFNmodels}

\setcounter{equation}{0}

In Ref. \cite{Kras2011} the following drift-free SFN models were identified; all 
are subcases of the Stephani \cite{Step1967,Kras1997} solution. 

\subsection{The spherically symmetric model}\label{sphsymm}

The metric of this model is
\begin{equation}\label{2.1}
{\rm d} s^2 = \left(\frac {F V,_t} V\right)^2 {\rm d} t^2 - \frac 1 {V^2}
\left({\rm d} r^2 + r^2 {\rm d} \vartheta^2 + r^2 \sin^2 \vartheta {\rm d}
\varphi^2\right),
\end{equation}
where $F(t)$ is an arbitrary function, related by $\theta = 3 / F$ to the
expansion scalar $\theta = {u^{\rho}};_{\rho}$ of the velocity field $u^{\alpha}
= \left(F V,_t\right)^{-1} V {\delta^{\alpha}}_0$. The function $V$ is
\begin{equation}\label{2.2}
V = B_1 + B_2 r^2 + \left(A_1 + A_2 r^2\right) S(t),
\end{equation}
where $(A_1, A_2, B_1, B_2)$ are arbitrary constants and $S(t)$ is an arbitrary
function. This model is conformally flat, but is more general than RW because
the pressure in it is spatially inhomogeneous. The RW limit results when
\begin{equation}\label{2.3}
A_1 \neq 0 \qquad {\rm and} \qquad B_2 = A_2 B_1 / A_1.
\end{equation}

\subsection{The plane symmetric model}\label{plansymm}

The metric is here
\begin{eqnarray}
{\rm d} s^2 &=& \left(\frac {F V,_t} V\right)^2 {\rm d} t^2 - \frac 1 {V^2}
\left({\rm d} x^2 + {\rm d} y^2 + {\rm d} z^2\right), \label{2.4} \\
V &=& B_1 + B_2 z + \left(A_1 + A_2 z\right) S(t), \label{2.5}
\end{eqnarray}
the meaning of all symbols being the same as before. Again, this model is
conformally flat, less general than Stephani \cite{Step1967,Kras1997}, but more
general than RW, and the prescription for the RW limit (here having $k \leq 0$
necessarily) is given by (\ref{2.3}). The resulting RW metric is represented in
untypical coordinates, see eqs. (7.8) in Ref. \cite{Kras2011} for a
transformation to a familiar form.

\subsection{The hyperbolically symmetric model}\label{hypsymm}

The metric is
\begin{equation}\label{2.6}
{\rm d} s^2 = \left(\frac {F V,_t} V\right)^2 {\rm d} t^2 - \frac 1 {V^2}
\left({\rm d} r^2 + {\rm d} \vartheta^2 + \sinh^2 \vartheta {\rm d}
\varphi^2\right)
\end{equation}
(note the missing factor $r^2$ compared to case A), where
\begin{equation}\label{2.7}
V = B_1 \sin r + B_2 \cos r + \left(A_1 \sin r + A_2 \cos r\right) S(t),
\end{equation}
the meaning of all symbols being again the same as in case A. As in both cases
above, this model is conformally flat, less general than Stephani
\cite{Step1967,Kras1997}, and more general than RW (this time with $k < 0$
necessarily). As in case B, the RW limit, resulting via (\ref{2.3}), is
represented in untypical coordinates; see Ref. \cite{Kras2011}.

\subsection{The axially symmetric model}\label{axisymm}

The general form of the metric is (\ref{2.4}), but this time
\begin{equation}\label{2.8}
V = {\displaystyle {\frac {C_5 - C_4 x_0 - \tfrac 1 2 {x_0}^2 + \tfrac 1 2
\left[\left(x - x_0\right)^2 + y^2 + z^2\right]} {D_1 x_0 + D_2}}},
\end{equation}
$(C_4, C_5, D_1, D_2)$ being arbitrary constants and $F(t)$, $x_0(t)$ being
arbitrary functions.\footnote{Equation (\ref{2.8}) corrects a typo in (A141) of
Ref. \cite{Kras2011}, where the whole term containing ${x_0}^2$ should be multiplied by 1/2. In (A140) of Ref. \cite{Kras2011} the second $(1/R),_{tt}$ should be $(1/R),_t$.} To calculate the RW limit, the following reparametrization is applied to (\ref{2.8}):
\begin{equation}\label{2.9}
x_0 = \delta U(t), \quad D_1 = d_1 / (\delta k), \quad D_2 = d_2 / k, \quad C_5
= 2 / k,
\end{equation}
where ($\delta, k)$ are constants. Then $\delta \to 0$ in (\ref{2.8}) gives
\begin{equation}\label{2.10}
V = {\displaystyle {\frac 2 {d_1 U + d_2} \ \left[1 + \tfrac 1 4 k \left(x^2 +
y^2 + z^2\right)\right]}},
\end{equation}
which clearly corresponds to the RW metric, the scale factor being $R(t) = 2 /
\left(d_1 U + d_2\right)$.

\section{The characteristic property of the drift-free
cases}\label{characteristic}

\setcounter{equation}{0}

In all the four cases presented in Sec. \ref{SFNmodels} the whole time
dependence is contained in $(1/V)^2$. Namely, in case A
\begin{eqnarray}\label{3.1}
{\rm d} s^2 &=& \frac 1 {V^2} \left\{\left[F \left(A_1 + A_2 r^2\right)
S,_t\right]^2 {\rm d} t^2 - {\rm d} r^2\right. \nonumber \\
&-& r^2 \left. \left({\rm d} \vartheta^2 + r^2 \sin^2 \vartheta {\rm d}
\varphi^2\right)\right\},
\end{eqnarray}
and the metric in braces is seen to be static. In cases B and C the factoring
out of time dependence occurs in similarly simple ways. In case D the
transformation
\begin{equation}\label{3.2}
t' = \int \frac {F x_{0,t}} {\left(D_1 x_0 + D_2\right)^2}\ {\rm d} t
\end{equation}
makes explicitly static the cofactor of $(1 / V)^2$. The RW models have the
same property.

In the general Stephani solution the conformal mapping to the Minkowski metric
involves mixing $t$ with spatial coordinates.\footnote{Because of the 5
arbitrary functions of $t$ in the Stephani metric, this conformal mapping cannot be calculated explicitly; we just know it exists, since the Weyl tensor is zero.} The congruence of curves in the Minkowski spacetime, to which the world lines of matter of a general Stephani solution are thereby mapped, must thus also display the non-RLP property.

In all cases listed in Sec. \ref{SFNmodels}, the time dependence factors out as
in (\ref{3.1}), and the world lines of cosmic medium are mapped into the world
lines of static observers. Relative to the congruence of static observers, all
light paths are evidently repeatable.

\section{Conclusion}\label{conclu}

\setcounter{equation}{0}

The result of Sec. \ref{characteristic} is the following

\medskip

{\bf Corollary 1}

In the Szekeres and SFN families of cosmological models, the subcases, in which
all null geodesics have repeatable paths, are characterized by the following
properties:

(1) The conformal curvature is zero.

(2) The time-dependence of the metric represented in comoving coordinates
factors out as in (\ref{3.1}).

In the Szekeres models, condition (1) is at the same time sufficient -- it
reduces the Szekeres models directly to the Friedmann limit.

\appendix

\section{An example of a timelike congruence in the Minkowski spacetime that
displays the non-RLP property}

The motivation for this example is explained in Sec. \ref{motiv}. Take the
Minkowski metric in the spherical coordinates
\begin{equation}\label{a.1}
{\rm d} s^2 = {\rm d} {t'}^2 - {\rm d} {r'}^2 - {r'}^2 \left({\rm d} \vartheta^2
+ \sin^2 \vartheta {\rm d} \varphi^2\right),
\end{equation}
and carry out the following transformation on it:
\begin{equation}\label{a.2}
t' = (r - t)^2 + 1 / (r + t)^2, \qquad r' = (r - t)^2 - 1 / (r + t)^2.
\end{equation}
The result is the metric
\begin{equation}\label{a.3}
{\rm d} s^2 = \frac 1 {(r + t)^4}\ {\rm d} \widetilde{s}^2,
\end{equation}
where
\begin{eqnarray}
&& {\rm d} \widetilde{s}^2 = 16 u \left({\rm d} t^2 - {\rm d} r^2\right) -
\left(u^2 - 1\right)^2 \left({\rm d} \vartheta^2 + \sin^2 \vartheta {\rm d}
\varphi^2\right), \nonumber \\
&& \label{a.4} \\
&& u \df r^2 - t^2. \label{a.5}
\end{eqnarray}
Now we assume that the curves with the unit tangent vector field $u^{\alpha} =
\left[(r + t)^2 / (4 \sqrt{u})\right] {\delta^{\alpha}}_0$ are world lines of
test observers and test light sources. Relative to this congruence, generic
light rays have nonrepeatable paths.

It suffices to consider (\ref{a.4}) with the unit tangent vector of the timelike congruence being $u^{\alpha} = [1 / (4 \sqrt{u})] {\delta^{\alpha}}_0$ instead
of (\ref{a.3}). Since conformal images of null geodesics are null geodesics, the only RLPs in (\ref{a.3}) will be the images of the RLPs in (\ref{a.4}), where
the mapping is defined by the same coordinates being used in both manifolds.

We investigate the conditions of repeatability by the method used in Refs.
\cite{KrBo2011,Kras2011}. We first observe that $r$ can be chosen as a
(nonaffine) parameter along open segments of null geodesics. Now consider two
light rays sent from the same source S at different instants toward the same
observer O. When the earlier ray arrives at a hypersurface $r = r_0$ at the
point with the coordinates $(t, \vartheta, \varphi)$, the later ray will arrive
at $r = r_0$ at the point $(t + \tau, \vartheta + \zeta, \varphi + \psi)$. The
equations of propagation of $(\tau, \zeta, \psi)$ are obtained from the geodesic equations by subtracting the equation for the earlier ray from the corresponding equation for the later ray, and linearizing the result in $(\tau, \zeta, \psi)$. The condition for a repeatable path is that $\zeta = \psi = 0$ is a solution of
the propagation equations.

Applying this operation and this condition to the geodesic equations
parametrized by $r$ we obtain
\begin{equation}\label{a.6}
\dr {\vartheta} r\ \chi = 0,
\end{equation}
where:
\begin{eqnarray}\label{a.7}
\chi &\df& \frac {3u^4 + 6u^2 - 1} {u \left(u^2 - 1\right)} t \tau \left[r
\left(\dr t r\right)^2 - 2 t \dr t r + r\right] \nonumber \\
&+& \left(3u^2 + 1\right) \left(r \dr t r\ \dr {\tau} r - \tau \dr t r - t \dr
{\tau} r\right).
\end{eqnarray}
One solution of (\ref{a.6}) is $\dril \vartheta r = 0$, which defines null
geodesics that are radial in the coordinates of (\ref{a.4}).

To find whether $\chi = 0$ has any solutions we proceed by the method described
in \cite{KrBo2011,Kras2011}. After a lengthy calculation (much simpler, though,
than in \cite{KrBo2011,Kras2011}) we obtain a contradiction, which means that no other RLPs than the radial null geodesics $\dril \vartheta r = \dril \varphi r = 0$ exist for ``comoving'' observers in the metric (\ref{a.4}). This implies

\medskip

{\bf Corollary 2}

With a suitably chosen timelike congruence of test observers and test light
sources, nonrepeatable light paths exist even in the Minkowski
spacetime.$\square$

\bigskip

{\bf Acknowledgments:} This work was supported by the Polish Ministry of Higher
Education grant N N202 104 838.

\end{document}